\documentclass[10pt]{iopart}
\usepackage{iopams}
\usepackage{color}
\usepackage{graphicx}

\begin{document}

\def\beq{\begin{eqnarray}}
\def\eeq{\end{eqnarray}}
\def\nn{{\nonumber}}

\title{Covariant approach of perturbations in Lovelock type brane 
gravity}

\author{Norma Bagatella-Flores,\,\,Cuauhtemoc Campuzano,\,\,Miguel Cruz 
and Efra\'\i n Rojas}

\address{Facultad de F\'\i sica, Universidad Veracruzana, 91000, Xalapa, 
Veracruz, M\'exico}

\eads{\mailto{nbagatella@uv.mx},\,\,\mailto{ccampuzano@uv.mx},\,\,
\mailto{miguelcruz02@uv.mx},\,\,
\mailto{efrojas@uv.mx}}

\begin{abstract}
We develop a covariant scheme to describe the dynamics of small perturbations 
on Lovelock type extended objects propagating in a flat Minkowski spacetime. The 
higher-dimensional analogue of the Jacobi equation in this theory becomes a wave 
type equation for a scalar field $\Phi$. Whithin this framework, we analyse the 
stability of membranes with a de Sitter geometry where we find that the Jacobi 
equation specializes to a Klein-Gordon (KG) equation for $\Phi$ possessing a 
tachyonic mass. This shows that, to some extent, these types of extended objects 
share the symmetries of the Dirac-Nambu-Goto (DNG) action which is by no means 
coincidental because the DNG model is the simplest included in this type of gravity.
\end{abstract}





\date{\today}

\section{Introduction}
\label{sec:intro}

Extended objects, sometimes referred to as membranes or branes for short, 
are surfaces of arbitrary dimension that are intended to represent many 
physical systems at diverse energy scales~\cite{carter92}. The most 
interesting action to describe a relativistic extended object propagating 
in a fixed background spacetime, based in the worldvolume geometry, is given 
by a local action involving a linear combination of higher-order curvature 
terms constructed by considering geometrical scalars~\cite{gregory,carter}. 
However, by appealing for both worldvolume reparametrization and background 
spacetime diffeomorphism invariances these geometrical scalars are limited. 
Among the plethora of worked geometric models, in first place we have the DNG 
model which is proportional to the volume swept out by the brane, the quadratic 
term in the mean extrinsic curvature introduced by Polyakov for a possible 
stringy description of QCD~\cite{polyakov86}, the Helfrich-Canham model in 
biophysics for describing fluid membranes~\cite{chinese99}, models for describing 
particle spinning particles either massive or masless~\cite{pisarski86,plyushchay89,nerse99}. 
Unfortunately, a clumsy fact is that the associated equations of motion (eom) 
are in general fourth-order equations in the field variables. 
Not all the worldvolume geometrical scalars fall into this category. For a given 
dimension $p$ there is a special subset of second-order terms which stabilize 
the extended object dynamics by maintaining the second-order eom thereby having 
potential physical applications. These are similar in form either to the original 
form of the Lovelock invariants in gravity or to their counterterms necessary 
in order to have a well-posed variational problem~\cite{lovelock,myers87,davis03,olea05,chilenos}.
This aspect makes a given theory, free from many of the pathologies that plague 
higher-order derivative theories (see also~\cite{zanelli} for a review on this topic). 
This fact is important because it assures no propagation of extra degrees of freedom. 
An appropriate description for extended objects realized in the form of a Born-Infeld 
type structure was developed in~\cite{cruzrojas2013} and named Lovelock type 
brane gravity. The interest in this subject has attracted quite a lot of attention 
recently by the attractive geometric structure and possible applications in 
accelerated cosmological scenarios~\cite{dgp,rham1,trodden1,trodden2}. 
In~\cite{cruzrojas2013}, the query on the existence and stability of the physical 
systems modeled by extended objects of this type was formulated so that here we 
enter the answer to the posed question.

The purpose of this paper is to provide a covariant framework to study the stability 
of small perturbations on Lovelock type extended objects moving in a flat Minkowski 
spacetime. The brane perturbations are guided by a Jacobi equation which is 
manifested through a wave type equation where the perturbations are described by a 
scalar field, $\Phi$, living in the $(p+1)$-dimensional worldvolume.  
This equation is characterized by a mass-like term expressed in terms of the 
worldvolume geometry. Because the mechanic content of an extended object, described by 
an action constructed from geometric scalars, is captured through their geometric degrees
of freedom (dof), it is expected that $\Phi$ will be in correspondence with the only 
dof that this type of models possesses. 

By specializing this framework to de Sitter branes with a spherical symmetry 
we are able to simplify the Jacobi equation to a KG one with a negative 
mass. This equation was extensively discussed by Garriga and Vilenkin in the case of 
DNG branes~\cite{vilenkin91,vilenkin92,vilenkin93}. We find that modes with 
angular momentum $L$ leading even eigenvalues $J$ associated to the spherical harmonics, 
produce an oscillatory behaviour that gets frozen at some time but subsequently restarts 
with some growing modes of oscillation which rapidly leading to instability thus ending 
up with highly nonspherical bubbles. For odd $J$ values the oscillatory behavior is 
drastically affected leading thus to instability more quickly. This result has been 
inherited for this extension to Lovelock type extended objects from the DNG model 
because, in some sense, the so-called Lovelock type brane gravity can be seen as a 
higher analogue of the DNG model with a extrinsic volume element~\cite{cruzrojas2013}. 
Therefore, it is all set to explore the physical implications arising from this 
perturbation analysis for this sort of gravity.

The paper is organized as follows. In Sec.~\ref{sec:2} we provide an overview 
of the Lovelock type brane gravity theory where we highlight the role that play 
some conserved tensors useful to study the mechanical content for this theory. 
We derive the first variation of the action governing the Lovelock type brane 
theory in Sec.~\ref{sec:3} in order to obtain the eom and to pave the way to carry 
out the analysis of perturbations. In Sec.~\ref{sec:4} we examine the second variation 
of the action and obtain the corresponding Jacobi equation which exhibits how the 
perturbations evolve. To illustrate our development, we specialize our framework to 
de Sitter membranes in Sec.~\ref{sec:5}. We conclude in Sec.~\ref{sec:6} with some 
comments and discuss our results. Appendix gathers useful information about the 
geometry of de Sitter branes floating in a Minkowski spacetime. For the sake of 
simplicity, we restrict our attention to closed extended objects with no physical 
boundaries.

\section{Lovelock type brane theory}
\label{sec:2}

Consider a dynamical extended object, $\Sigma$, of dimension $p$ moving in a 
$N=(p+2)$-dimensional Minkowski spacetime ${\cal M}$ with metric $\eta_{\mu \nu}$ 
$(\mu,\nu =0,1,\ldots, p+1)$. The worldvolume $m$, is an oriented  timelike hypersurface 
manifold of dimension $p+1$, described by the embedding functions $y^\mu = X^\mu(x^a)$ 
where $y^\mu$ are local coordinates of ${\cal M}$ and  $x^a$ are local coordinates 
of $m$, and $X^\mu$ are the embedding functions $(a,b= 0,1,\ldots,p)$. The most 
important derivatives of $X^\mu$ enter the game through the induced metric tensor 
$g_{ab} = \eta_{\mu \nu} e^\mu{}_a e^\nu{}_b:= e_a \cdot e_b$ and the extrinsic 
curvature $K_{ab} = - n \cdot \nabla_a e_b = K_{ba}$ where $e^\mu{}_a = \partial_a 
X^\mu$ are the tangent vectors to $m$, $n^\mu$ is the spacelike unit normal 
vector to $m$, and $\nabla_a$ is the covariant derivative compatible with $g_{ab}$.

For a $(p+1)$-dimensional worldvolume described by the variables $X^\mu$ 
the action~\cite{cruzrojas2013}
\begin{equation}
S[X] =  \int_m d^{p+1} x \, \sqrt{-g} \sum_{n=0} ^{p+1} \alpha_n\,L_n (g_{ab},
K_{ab}),
\label{eq:Lbaction}
\end{equation}
where
\begin{equation}
L_n (g_{ab}, K_{ab}) = \delta^{a_1 a_2 a_3 \cdots a_n} _{b_1 b_2 b_3 \cdots b_n}
 K^{b_1}{}_{a_1} K^{b_2}{}_{a_2} K^{b_3}{}_{a_3} \cdots K^{b_n}{}_{a_n},
\label{eq:lovelock-brane}
\end{equation}
ensures that the field equations of motion are of second order. 
Here, $\delta^{a_1 a_2 a_3 \ldots a_n} _{b_1 b_2 b_3 
\ldots b_n}$ denotes the generalized Kronecker delta (gKd), 
$g= \textrm{det} (g_{ab})$ and $\alpha_n$ are constants with appropriate dimensions. 
We set $L_0 = 1$. This action is invariant under reparametrizations of 
the worldvolume. The Lagrangian (\ref{eq:lovelock-brane}) is a polynomial 
of degree $n \leq p+1$ in the extrinsic curvature so that the action~(\ref{eq:Lbaction}) 
is a second-order derivative theory. The geometrical invariants~(\ref{eq:lovelock-brane}) 
are to be referred to as {\it Lovelock type brane invariants} (LBI). By construction, 
these scalars vanish for $n>p+1$ whereas the term with $n= p+1$ 
corresponds to a topological invariant not contributing to the field equations. 
Since $X^\mu$ are the independent variables instead of the metric, we then have one 
greater number of Lovelock type terms contrary to the pure gravity case. We can find 
a complete list of the first LBI in~\cite{cruzrojas2013}.

For even $n$ we recognize the form of the Gauss-Bonnet (GB) terms but expressed 
now in terms of the worldvolume geometry. For example, for $n=0$ we have the DNG 
Lagrangian, for $n=2$ we have the Regge-Teitelboim (RT) 
model~\cite{rt,tapia89,davidson1,davidson2,pavsic2002,ostrogradski,modified2012}, for $n=4$ we have 
the form of the GB Lagrangian which for $p>3$ produces non-vanishing eom
with ghost-free contribution. 
On the other side, for odd $n$ the corresponding Lagrangians look like the 
Gibbons-Hawking-York boundary terms which may exist if we have the presence of 
bulk Lovelock invariants~(see~\cite{lovelock,cruzrojas2013} for details).

Some remarks are in order. To avoid ambiguities for possible gauge 
invariance for the case of odd $n$ Lagrangians, 
we assume that $n^\mu$ is such that it is pointing outward to $m$. 
Furthermore, one should not get confused and think that these LBI come from 
the original pure Lovelock theory as counter-terms; as discussed in~\cite{carter92,defo}, 
such geometrical scalars~(\ref{eq:lovelock-brane}) generate dynamics 
without the need to be considered as surface terms, where the only 
relevant degrees of freedom are the ones associated with the geometric configuration 
of the worldvolume itself. Moreover, observe that 
definition~(\ref{eq:lovelock-brane}) coincides with the expression of 
the determinant of $K^a{}_b$. Indeed, the functional
\begin{equation}
S = \int_m d^{p+1} x \, \sqrt{-g}\, \mbox{det}\,(K^a{}_b), 
\end{equation}
correspond to the Gauss-Bonnet topological invariant which is a conformal
invariant functional with respect to conformal transformations of the 
worldvolume geometry~\cite{conformal96}. Hereafter, as a simplification in 
the notation, the differential $d^{p+1}x$ wherever a worldvolume integration 
is performed will be absorbed.

By virtue of the properties of the gKd function one can define the important 
tensors
\begin{equation}
\label{eq:conserved1}
J^{a}_{(n)b} := \delta^{a a_1 a_2 a_3 \ldots a_n} _{b b_1 b_2 b_3 \ldots b_n}
K^{b_1}{}_{a_1} K^{b_2}{}_{a_2} K^{b_3}{}_{a_3} \ldots K^{b_n}{}_{a_n}.
\end{equation}
These are symmetric and conserved because $\nabla_a J^{ab} _{(n)} = 0$. This fact 
is shown by using the properties of the gKd and the Codazzi-Mainardi 
integrability condition for surfaces of arbitrary dimension when the ambient
spacetime is Minkowski, $\nabla_a K_{bc} = \nabla_b K_{ac}$. Notice 
that, for a $(p+1)$-dimensional worldvolume there are at most an equal number of 
conserved tensors $J^{ab} _{(n)}$. As showed in~\cite{cruzrojas2013}, by expanding 
out the determinant involved in~(\ref{eq:conserved1}) in terms of minors we obtain 
a useful recursion relation
\begin{equation}
 J^a _{(n) b} = \delta^a _b \,L_n - n K^a{}_c J^{c}_{(n-1)b}.
\label{eq:identity}
\end{equation}
As before, in~\cite{cruzrojas2013} we have a detailed list of the 
first conserved Lovelock tensors.
The $J^{ab} _{(n)}$ are to be referred to as {\it Lovelock brane tensors} 
(LBT). We already have some familiarity with the conservation property of 
(\ref{eq:conserved1}). Indeed, $J_{(0)} ^{ab}= g^{ab}$ is conserved because 
we have a Levi-Civita connection; $J_{(1)} ^{ab} = g^{ab} K - K^{ab}$ is 
conserved due to the contraction of the Codazzi-Mainardi integrability 
condition whereas $J_{(2)} ^{ab} = {\cal R} g^{ab} - 2 {\cal R}^{ab}$ is nothing 
but the worldvolume Einstein tensor which is conserved by the Bianchi identity. 

The contraction of Eq.~(\ref{eq:conserved1}) with the extrinsic curvature tensor, 
by considering Eq.~(\ref{eq:lovelock-brane}), provides an identity among the LBT 
and the LBI 
\begin{equation}
J^{ab} _{(n)} K_{ab} = L_{n+1}.
\label{eq:relation}
\end{equation}

\section{Extremality}
\label{sec:3}

The main fact behind the Lovelock brane invariants is that their associated 
eom are of second-order in the derivatives of the dynamical variables $X^\mu$. 
To prove this we consider directly the variation of the action~(\ref{eq:Lbaction})  
\begin{equation}
\delta S = \int_m \left[ \delta \left( \sqrt{-g}\right) 
\sum_{n=0} ^{p+1} \alpha_n L_n + \sqrt{-g} \sum_{n=0} ^{p+1} \alpha_n 
\delta L_n \right].
\label{var1}
\end{equation}
Carrying out the variations under the integral sign we have first that $\delta 
\sqrt{-g} = (\sqrt{-g}/2) g^{ab}\delta g_{ab}$. With regards the second one we 
have from~(\ref{eq:lovelock-brane}) that
\begin{eqnarray}
\delta L_n &=& n \delta^{a_1 a_2 a_3\cdots a_n} _{b_1 b_2 b_3 \cdots b_n} 
K^{b_2}{}_{a_2} K^{b_3}{}_{a_3} \cdots K^{b_n}{}_{a_n} \delta  K^{b_1}{}_{a_1},
\nonumber
\\
&=& n J^a _{(n-1)b} \delta  K^{b}{}_{a},
\label{var2}
\end{eqnarray}
where we have used the antisymmetric properties of the gKd and the 
definition~(\ref{eq:conserved1}). We also require that $\delta K^a{}_b = 
\delta (g^{ac}K_{cb}) = - g^{ad} K^c{}_b \delta g_{dc} + g^{ac} \delta K_{cb}$. 
Thus, the variation~(\ref{var2}) becomes
\begin{eqnarray}
\delta L_n &=& \left( J^{ab} _{(n)} - g^{ab} L_n \right)\delta g_{ab} + n J^{ab} _{(n-1)} 
\delta K_{ab},
\label{var3}
\end{eqnarray}
where we have used the identity~(\ref{eq:identity}). Relation~(\ref{var3}) allows 
us to write the variation~(\ref{var1}) in the form
\begin{equation}
\delta S = \int_m \sqrt{-g} \sum_{n=0} ^{p+1} \alpha_n \left[ 
\left( J^{ab} _{(n)} - \frac{1}{2} g^{ab} L_n \right) \delta g_{ab}
+ n J^{ab} _{(n-1)} \delta K_{ab}
\right].
\label{var4}
\end{equation}
At this stage what we have is the response of the action~(\ref{eq:Lbaction}) 
to small changes in the worldvolume 
\begin{equation}
X^\mu (x^a) \rightarrow X^\mu (x^a) + \delta X^\mu (x^a),
\label{eq:defo}
\end{equation}
through the variations $\delta g_{ab}$ and $\delta K_{ab}$.
Now, it should be emphasized that only motions tranverse to the worldvolume are 
physically observable so that we shall consider only $\delta_\perp X^\mu = \Phi \,n^\mu$
where $\Phi$ is assumed to be small but is otherwise an arbitrary function of $x^a$.
This is reflected in the fact that the most important variations to be considered 
are~\cite{defo}
\begin{eqnarray}
\delta_\perp g_{ab} &=& 2 K_{ab}\,\Phi,
\label{eq:var-g}
\\
\delta_\perp K_{ab} &=& - \nabla_a \nabla_b \Phi + K_{ac} K^c{}_b\,\Phi.
\label{eq:var-K}
\end{eqnarray}
If we replace these expressions into~(\ref{var4}) we get
\begin{eqnarray}
\delta S &=& \int_m \sqrt{-g} \sum_{n=0} ^{p+1} \alpha_n\left( 
2 J^{ab} _{(n)} K_{ab} - K L_n  + n K^c{}_a J^{ab} _{(n-1)} K_{cb} 
\right) \Phi 
\nonumber
\\
&+&  \int_m  \nabla_a \left( - \sqrt{-g}\sum_{n=0} ^{p+1} \alpha_n\,
n J^{ab} _{(n-1)} \nabla_b \Phi \right),
\nonumber  
\end{eqnarray}
where we have used the conservation of the LBT. By using the identities~(\ref{eq:identity})
and (\ref{eq:relation}) we finally obtain the variation of the action $S$ in the form
\begin{equation}
\delta S = \int_m \sqrt{-g} \,{\cal E} \Phi 
+ \int_m \sqrt{-g}\,\nabla_a Q^a [\Phi],
\end{equation}
where
\begin{equation}
{\cal E} = \sum_{n=0} ^{p+1}\alpha_n 
J^{ab} _{(n)} K_{ab} = 0 ,
\label{eq:eom1}
\end{equation}
is the Euler-Lagrange (EL) equation of motion which in turn can be expressed
in terms of the LBI through the identity~(\ref{eq:relation}). In addition,
\begin{equation}
Q^a := \sum_{n=0} ^{p+1} \alpha_n Q^a _{(n)}
= - \sum_{n=1} ^{p+1} \alpha_n \,n J^{ab} _{(n-1)} \nabla_b ,
\end{equation}
is a differential operator defined on the worldvolume. Obviously, we have only 
one equation of motion which is second-order in the field 
variables. This fact, in particular, means that we have only one physical degree of 
freedom for this type of branes. Indeed, the physically observable measure of the 
deformation of the surface is the breathing mode provided by the scalar field $\Phi$.
In summary, under the deformation~(\ref{eq:defo}), the first variation of the
action~(\ref{eq:Lbaction}) yields
\begin{equation}
\delta S = \int_m \sqrt{-g} \sum_{n=0} ^{p+1} \alpha_n\,J^{ab} _{(n)} K_{ab}\,
\Phi = 0,
\end{equation}
where~(\ref{eq:eom1}) is the result of the physical transverse motions. As expected,
we may think of the eom~(\ref{eq:eom1}) as the generalization  of the condition for
extremal hypersurfaces in the sense that we have the vanishing of the trace of the 
extrinsic curvature where the conserved tensors $J^{ab} _{(n)}$ play the role of a metric.

\section{Linearized normal perturbations on the worldvolume}
\label{sec:4}

We turn now to compute the second variation of the action $S$. 
Unlike the first variation to obtain the equation of motion, 
in general in order to have a well posed second variation it is not allowed to neglect 
the tangential deformations of the surface. This is related to the fact that a finite 
tangential deformation, unlike its infinitesimal counterpart, is not a simple 
reparametrization of the surface~\cite{defo2}. From the second variation
\begin{equation}
\delta^2 S = \int_m \sum_{n=0} ^{p+1} \alpha_n \delta \left( \sqrt{-g} 
\,J^{ab} _{(n)} K_{ab}\,\Phi  \right),
\label{eq:secV}
\end{equation}
clearly we have that the relevant equation is
\begin{equation}
 \sum_{n=0} ^{p+1} \alpha_n \delta \left( J^{ab} _{(n)} K_{ab} \right) = 0.
\label{eq:eom2}
\end{equation}
It has been formally proved in~\cite{defo2} that if ${\cal E}$ 
represents the EL equation of motion, then the total variation $\delta \left( 
\sqrt{-g}\, {\cal E}\,\Phi \right)= \delta_\perp \left( \sqrt{-g}\, {\cal E}
\right)\Phi = {\cal L} \Phi$ where ${\cal L}$ is a local differential operator
so that it is only necessary to consider the variations~(\ref{eq:var-g}) 
and~(\ref{eq:var-K}). Accordingly, to evaluate~(\ref{eq:eom2}) we begin first 
by using the definition~(\ref{eq:conserved1}) to obtain
\begin{equation}
\delta_\perp J^{ab} _{(n)} 
= \delta_\perp g^{bc} \,J^a _{(n)c} 
+ n  g^{bc} \,\delta^{a a_1 a_2 \cdots a_n} _{c b_1 b_2 \cdots b_n}K^{b_2}{}_{a_2} 
\cdots K^{b_n}{}_{a_n} \,\delta_\perp K^{b_1}{}_{a_1}.
\nonumber
\end{equation}
Now, as a result of the contraction of this relation with the extrinsic curvature 
we have
\begin{eqnarray}
\fl
\hspace{2cm} K_{ab} \delta_\perp J^{ab} _{(n)}  =
\delta_\perp g^{bc} J^a _{(n)c}K_{ab} 
+ n\,\delta^{ a_1 a a_2 \cdots a_n} _{b_1 c b_2 \cdots b_n}
K^c{}_a K^{b_2}{}_{a_2} 
\cdots K^{b_n}{}_{a_n} \delta_\perp K^{b_1}{}_{a_1},
\nonumber
\\
\hspace{1.15cm} =  \delta_\perp g^{bc} J^a _{(n)c}K_{ab} + nJ^a _{(n)b}\delta_\perp K^b{}_a,
\nonumber
\end{eqnarray}
where in the first line of the right-hand side we have used, once again, the 
antisymmetric properties of the gKd. Further, observe that $\delta_\perp K^b{}_a 
= \delta_\perp g^{bc} K_{ca} + g^{bc} \delta_\perp K_{c a}$ and therefore
\begin{equation}
\nonumber
K_{ab} \delta_\perp J^{ab} _{(n)} 
= (n+1) J^a _{(n)c} K_{ab} \delta_\perp g^{bc} + n J^{ab} _{(n)} \delta_\perp K_{ab}.
\end{equation}
In this manner, from~(\ref{eq:eom2}) we find that
\begin{equation}
\delta_\perp \left( J^{ab} _{(n)} K_{ab} \right) = (n+1) J^a _{(n)c} K_{ab} 
\delta_\perp g^{bc} + (n+1) J^{ab} _{(n)} \delta_\perp K_{ab}.
\label{eq:expre}
\end{equation}
As mentioned before, only motions transverse to the worldvolume are physically 
observable so, from~(\ref{eq:var-g}) one has $\delta_\perp g^{ab} = -g^{ac}g^{bd} 
\delta_\perp g_{cd} = - 2 K^{ab} \Phi$. Hence, by plugging this variation 
and~(\ref{eq:var-K}) into expression~(\ref{eq:expre}) we get
 \begin{eqnarray}
\delta \left( J^{ab} _{(n)} K_{ab} \right) 
&=& - (n+1) \left[J^{ab} _{(n)} \nabla_a \nabla_b \Phi 
+ J^{ab} _{(n)} K_a{}^c K_{bc}\,\Phi \right] .
\nonumber
\end{eqnarray}
It therefore follows that when we treat the second variation~(\ref{eq:secV}) 
as an action, $S':=\delta^2 S$, we find
\begin{equation}
S' = - \int_m \sqrt{-g} \sum_{n=0} ^{p+1} \alpha_n  
\,(n+1) \Phi \left[J^{ab} _{(n)} 
\nabla_a \nabla_b \Phi + J^{ab} _{(n)} K_a{}^c K_{bc}\,\Phi \right].
\label{eq:secV2}
\end{equation}
The variation of this action with respect to $\Phi$ leads immediately to 
the expression
$\sum_{n=0} ^{p+1} \alpha_n (J^{ab} _{(n)} \nabla_a \nabla_b \Phi 
+  J^{ab} _{(n)} K_a{}^c K_{bc}\,\Phi) = 0$, or more suggestively as
\begin{equation}
\sum_{n=0} ^{p+1} \alpha_n (n+1) \left[ J^{ab} _{(n)} \nabla_a \nabla_b \Phi 
+  M_{(n)} ^2\,\Phi \right] = 0,
\label{eq:ray1}
\end{equation}
where 
\begin{equation}
 M^2 _{(n)} :=  J^{ab} _{(n)} K_{a}{}^c K_{bc},
 \label{eq:M}
\end{equation}
is a purely geometric quantity that plays the role of a mass-like term. 
The equation~(\ref{eq:ray1}) corresponds to the \textit{Jacobi equation} in the 
Lovelock type brane gravity. On geometrical grounds, the Jacobi equation describes 
the behavior of Lovelock type surfaces which are close or in the neighborhood 
of a reference surface one. Clearly, we may think of this equation as the second-order 
generalization of the corresponding Jacobi equation for the case of DNG branes so 
that we provide a generalization of the wave equation developed for an arbitrary DNG 
extended object moving in a Minkowski 
spacetime~\cite{vilenkin91,vilenkin92,vilenkin93,defo,guven93,sayan06,raycha95}.
Consequently, we will regard~(\ref{eq:ray1}) as the equation of motion for the small 
perturbations $\Phi$. Thus, the second variation of~(\ref{eq:Lbaction}) can be expressed as
\begin{equation}
\delta^2 S [X^\mu] = \int_m  \sqrt{-g}\,\Phi\,{\cal L} \Phi,
\end{equation}
where
\begin{equation}
{\cal L} = \sum_{n=0} ^{p+1} \alpha_n {\cal L}_{(n)} =  \sum_{n=0} ^{p+1} \alpha_n (n+1) \left[
J^{ab} _{(n)} \nabla_a \nabla_b + M^2 _{(n)} \right].
\end{equation}
Moreover, taking advantage of the conservation property of the LBT,
a convenient way to express Eq.~(\ref{eq:ray1}) is as follows
\begin{equation}
\sum_{n=0} ^{p+1} \alpha_n (n+1) \left[ \frac{1}{\sqrt{-g}} \partial_a \left( \sqrt{-g} 
J^{ab} _{(n)} \partial_b \Phi \right) + M^2 _{(n)} \,\Phi \right] = 0.
\label{eq:wave2}
\end{equation} 
The solutions of the Jacobi equations~(\ref{eq:ray1}) address the question of stability 
through the nature of the modes $\Phi$. If we find that the resulting modes are oscillatory, 
then they indicate stability of the system whereas growing modes confirm the existence of 
an instability. In this sense, the only degree of freedom describing the deformation
of any Lovelock type extended object is the breathing mode of the worldvolume. It should 
be mentioned that we may arrive at the same Eq.~(\ref{eq:ray1}) in a somewhat different 
manner. That is, from a non-perturbative framework by considering a purely kinematical 
description of the normal deformations of the worldvolume~\cite{raycha95}. There, the 
Jacobi equation corresponds to a higher-dimensional analogue of the Raychaudhuri equation 
describing point particles.   

From the Lovelock brane tensors~(\ref{eq:conserved1}) and (\ref{eq:M})
we have at few order that
\begin{eqnarray}
M_{(0)} ^2 &=& K_{ab} K^{ab},
\\
M_{(1)} ^2 & = &  K^{ab} {\cal R}_{ab},
\\
M_{(2)} ^2 & = &
{\cal R} K_{ab} K^{ab} - 2 {\cal R}^b{}_c K^c{}_a K^a{}_b,
\end{eqnarray}
so that the wave-like equation can be rewritten explicitly. For example, for $n=0$, 
we know that $J^{ab} _{(0)} = g^{ab}$. Thereby, Eq.~(\ref{eq:wave2}) becomes
\begin{equation}
\nonumber
\Box \Phi + K_{ab} K^{ab} \Phi = 0,
\end{equation}
where $\Box = g^{ab}\nabla_a \nabla_b$ stands for the d'Alembertian operator defined
on $m$. This case has been studied exhaustively in many 
contributions~\cite{vilenkin91,vilenkin92,vilenkin93,guven93,larsen94}.

\section{Perturbations on de Sitter membranes}
\label{sec:5}

We now confine our attention to a specific geometry. Suppose that $\Sigma$ is a 
$p$-dimensional de Sitter membrane moving in a Minkowski spacetime described by 
the action~(\ref{eq:Lbaction}). The metric on $m$ can be expressed in the form
\begin{equation}
dS_{p+1} ^2 = - d\tau^2 + H^{-2} \cosh^2 (H\tau) d\Omega_{(p)} ^2, 
\end{equation}
where $d \Omega_{(p)} ^2 = d\theta_1 ^2 + \sin^2 \theta_1 d\theta_2 ^2
+ \sin^2 \theta_1 \sin^2 \theta_2 d\theta_3 ^2 + \ldots + \sin^2 \theta_1 \sin^2 \theta_2 
\cdots \sin^2 \theta_{p-1} d\theta_p$ is the metric on the 
$p$-sphere and $H$ is a constant (see~\ref{app1} for details). 
Thus, $\sqrt{-g} = H^{-p} \cosh^p (H\tau) \prod_{r=1} ^p \sin^{p-r} \theta_r$. 
Being a maximally symmetric spacetime, the Lovelock type tensors as well as 
the Lovelock type invariants become 
\begin{eqnarray}
J^{ab} _{(n)} &=& C_{(p,n)} H^n g^{ab}, 
\label{eq:tensor1}
\\
L_n &=& C_{(p+1,n)}H^n,
\label{eq:tensor2}
\end{eqnarray}
where we have introduced the notation $C_{(p,n)} = \Gamma (p+1) / \Gamma (p-n+1)$ 
and $\Gamma (n)$ is the Gamma function. The expression~(\ref{eq:tensor1}), 
together with the relation~(\ref{eq:A8}), when substituted into the expression 
for the mass-like term~(\ref{eq:M}) lead to
\begin{equation}
M_{(n)} ^2 =  (p+1) C_{(p,n)} H^{n+2}.
\label{eq:M2}
\end{equation}
It is evident now that the quantities~(\ref{eq:tensor1}) 
and~(\ref{eq:M2}), when inserted into the relation~(\ref{eq:wave2}), give rise
\begin{equation}
\sum_{n=0} ^{p+1} \alpha_n (n+1) H^n C_{(p,n)} \left[ \Box \Phi 
+ (p+1) H^2 \Phi \right] = 0,
\label{eq:wave33}
\end{equation}
which in turn implies that
\begin{equation}
 \Box \Phi + (p+1) H^2 \Phi  = 0.
\label{eq:wave3}
\end{equation}
Therefore, the theory of perturbations on de Sitter Lovelock type membranes
simplifies enormously to the analysis of a Klein-Gordon equation for a scalar 
field $\Phi$ with a negative mass $m^2 := - (p+1)H^2$, as in the case of DNG 
extended objects~\cite{vilenkin91,guven93,sayan06}. To understand this point 
we only insert the emergent expression for the Lagrangians~(\ref{eq:tensor2}) 
into the action~(\ref{eq:Lbaction}) and observe that a DNG action emerges.

To solve Eq.~(\ref{eq:wave3}) it is convenient to consider a decomposition
into spherical modes
\begin{equation}
\fl
\Phi (\tau,\theta_i) = \sum_{LM} \phi_L (\tau) Y_{LM} (\theta_i).
\qquad \quad
\begin{array}{ll}
L &= 0,1,\ldots, \infty.
\\
M &= -L, -L +1, \ldots, 0 , \ldots, L-1, L.
\\
i &= 1,2,\ldots,p.
\end{array}
\label{eq:ansatz}
\end{equation}
Here, $Y_{LM}$ stands for the spherical harmonics on the $p$-sphere
satisfying $\Delta Y_{LM} = -J \,Y_{LM}$ where $\Delta$ is the Laplacian 
operator defined on the unit $p$-sphere and $J$ denotes the eigenvalues 
of the spherical harmonics given by $J = L (L + p - 1)$. 
The solution $\phi_L (\tau)$ is given by a linear combination of the 
associated Legendre functions $P^\mu _\nu ( \tanh (H\tau))$ and  
$Q^\mu _\nu (\tanh (H\tau))$ (see \cite{vilenkin91}) for details). 

\begin{figure}[htb!]
\centering
\includegraphics[angle=0,width=5.5cm,height=5cm]{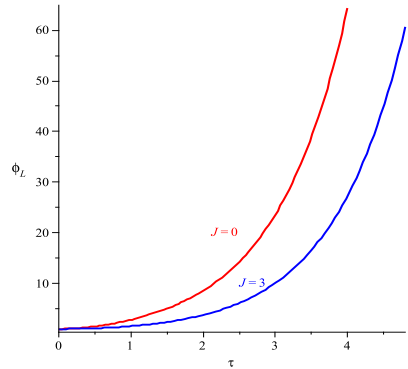}
\caption{The bubble behaviour for $J=0$ and $J=3$ values.
These are not perturbations but time translation and spatial rotation of
the bubble for the lowest modes $L=0$ and $L=1$, respectively.}
\label{fig:1}
\end{figure}

\begin{figure}[htbp!]
\centering
\includegraphics[width=5.5cm,height=4.5cm]{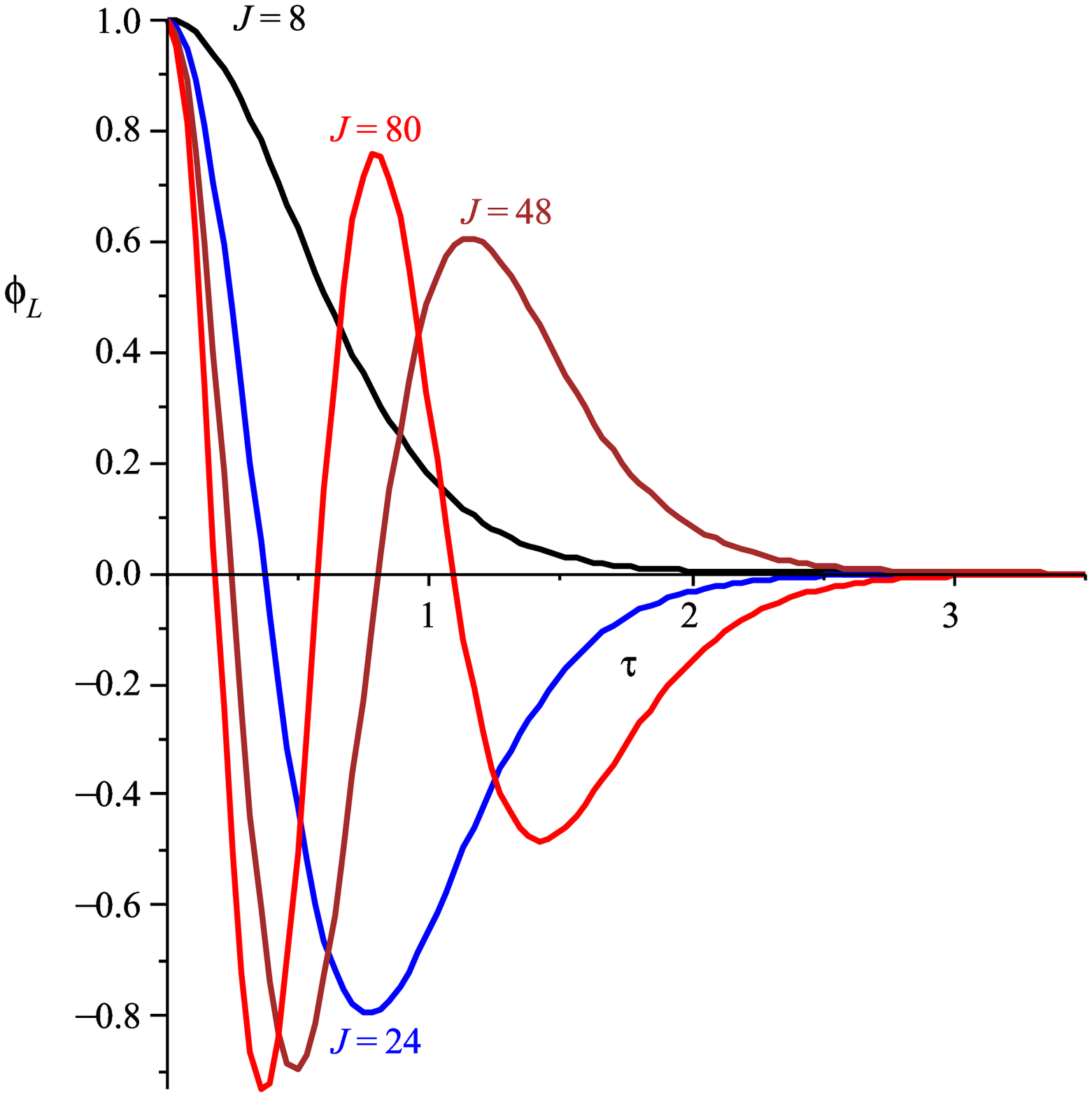}
\includegraphics[width=5.5cm,height=4.5cm]{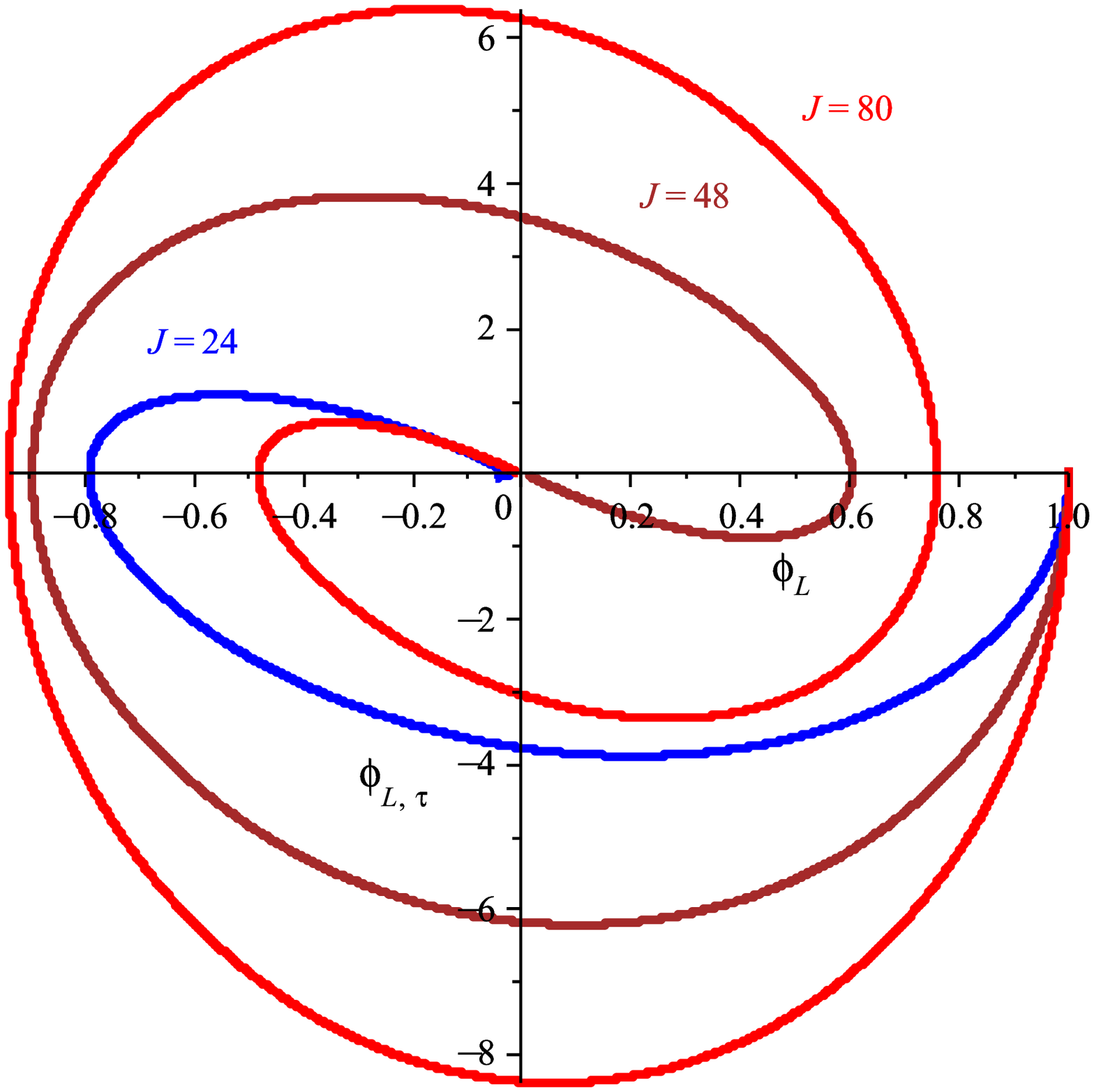}
\includegraphics[width=5.5cm,height=4.5cm]{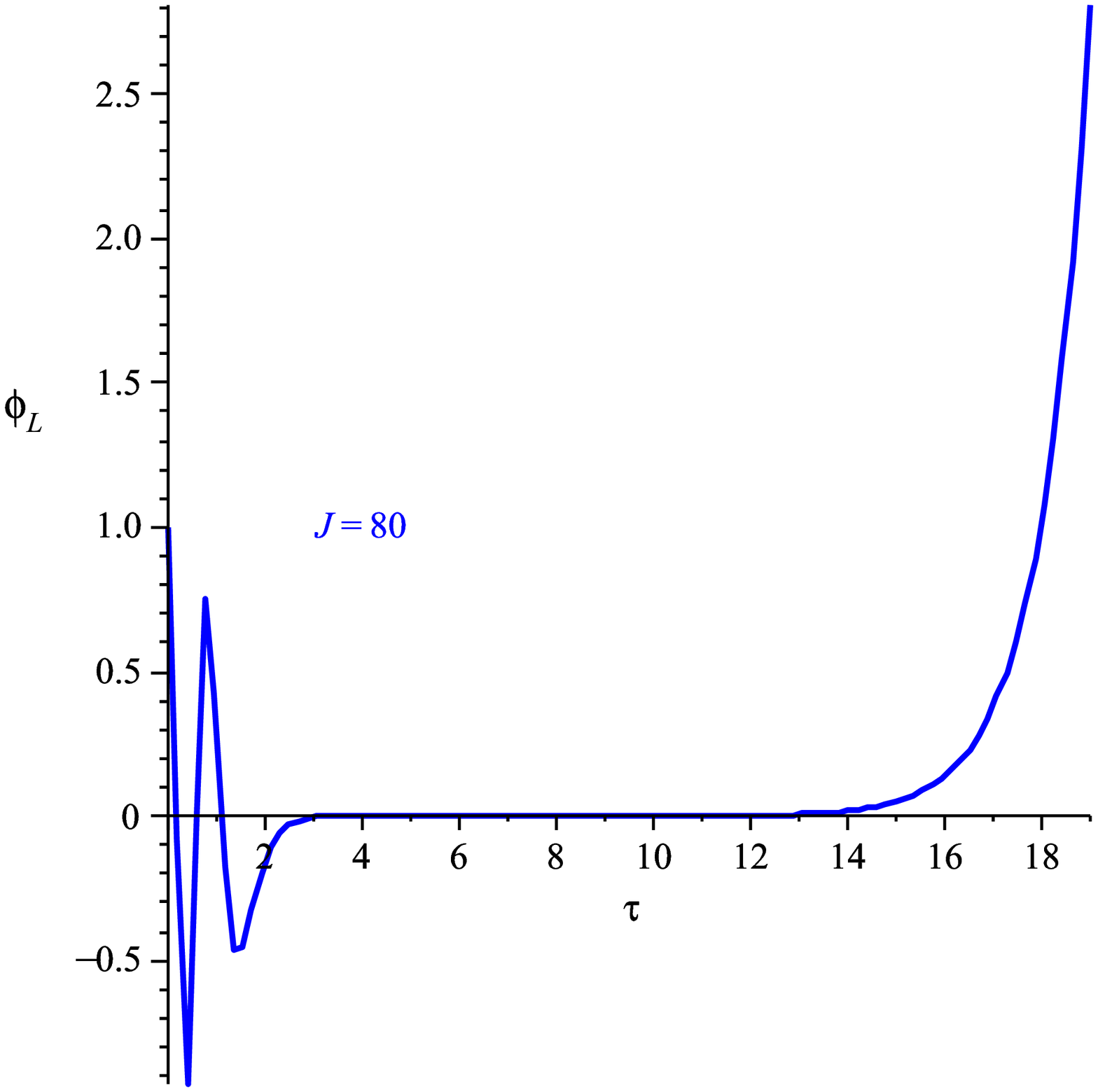}
\includegraphics[width=5.5cm,height=4.5cm]{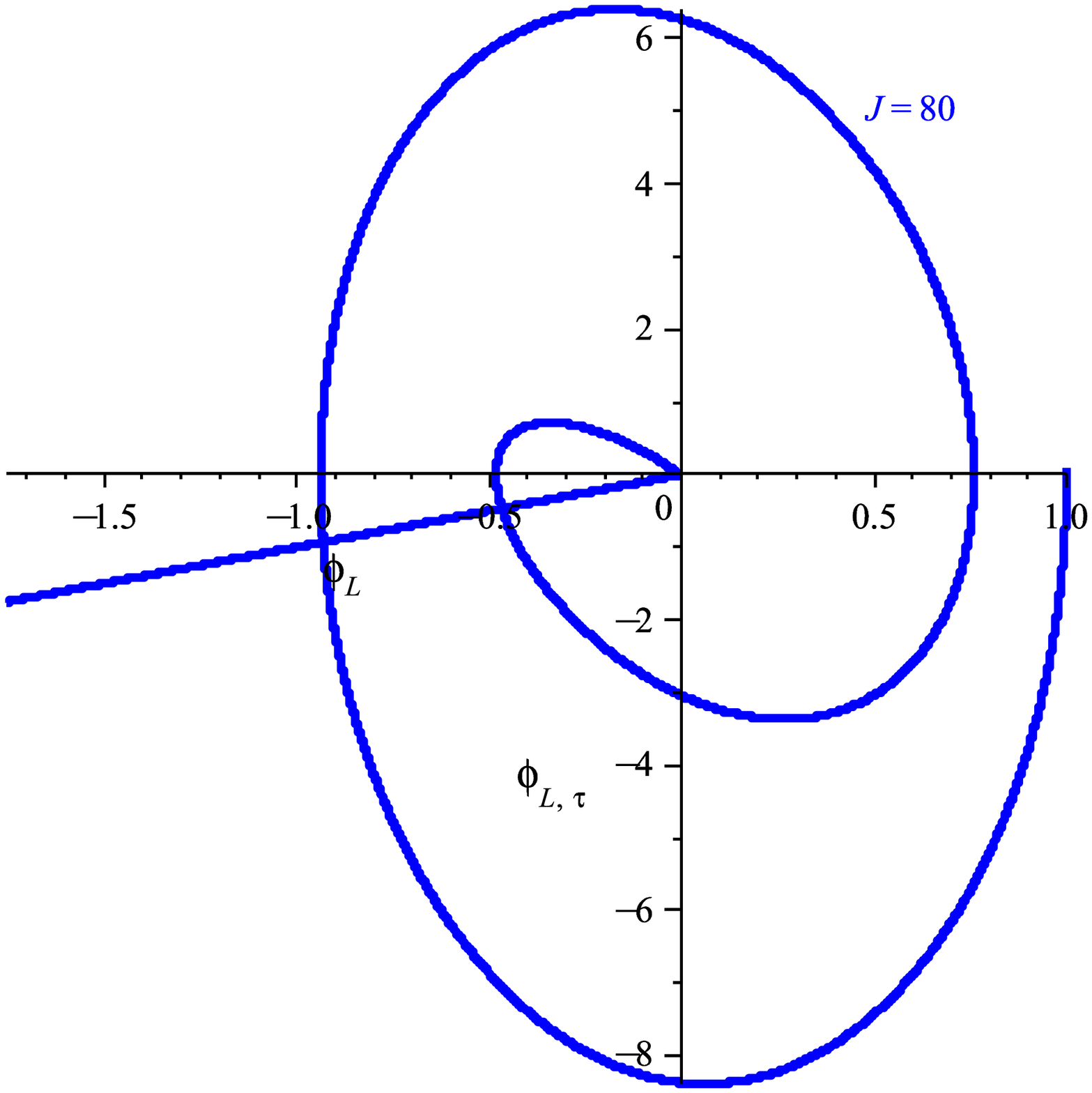}
\caption{Oscillatory behaviour obtained for the even values of the parameter $J$
considering short and large times. The left pictures are depicted in the configuration 
space while on the right were plotted in the corresponding phase space.}
\label{fig:2}
\end{figure}

\begin{figure}[htb!]
\centering
\includegraphics[width=5.5cm,height=4.5cm]{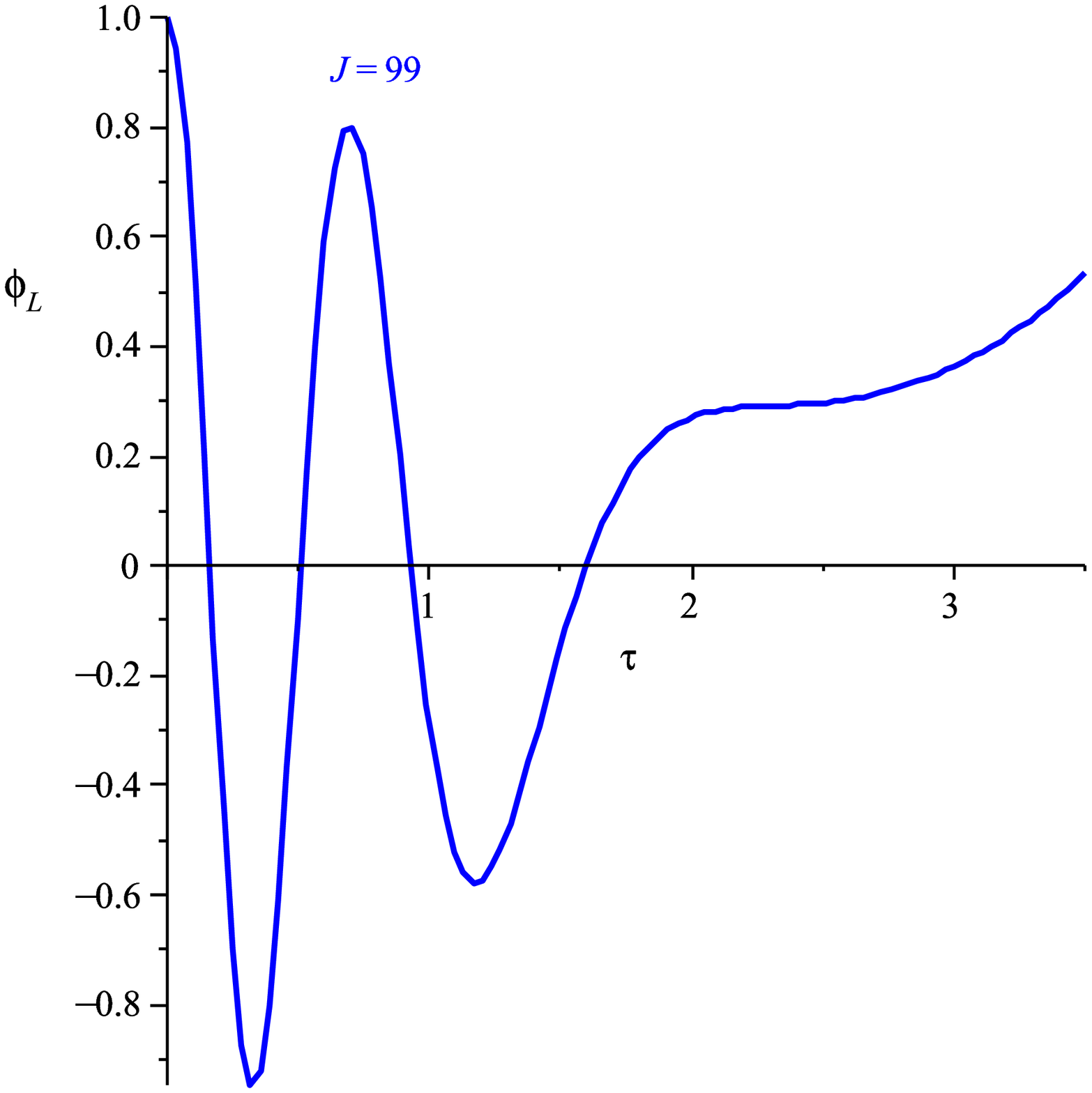}
\includegraphics[width=5.5cm,height=4.5cm]{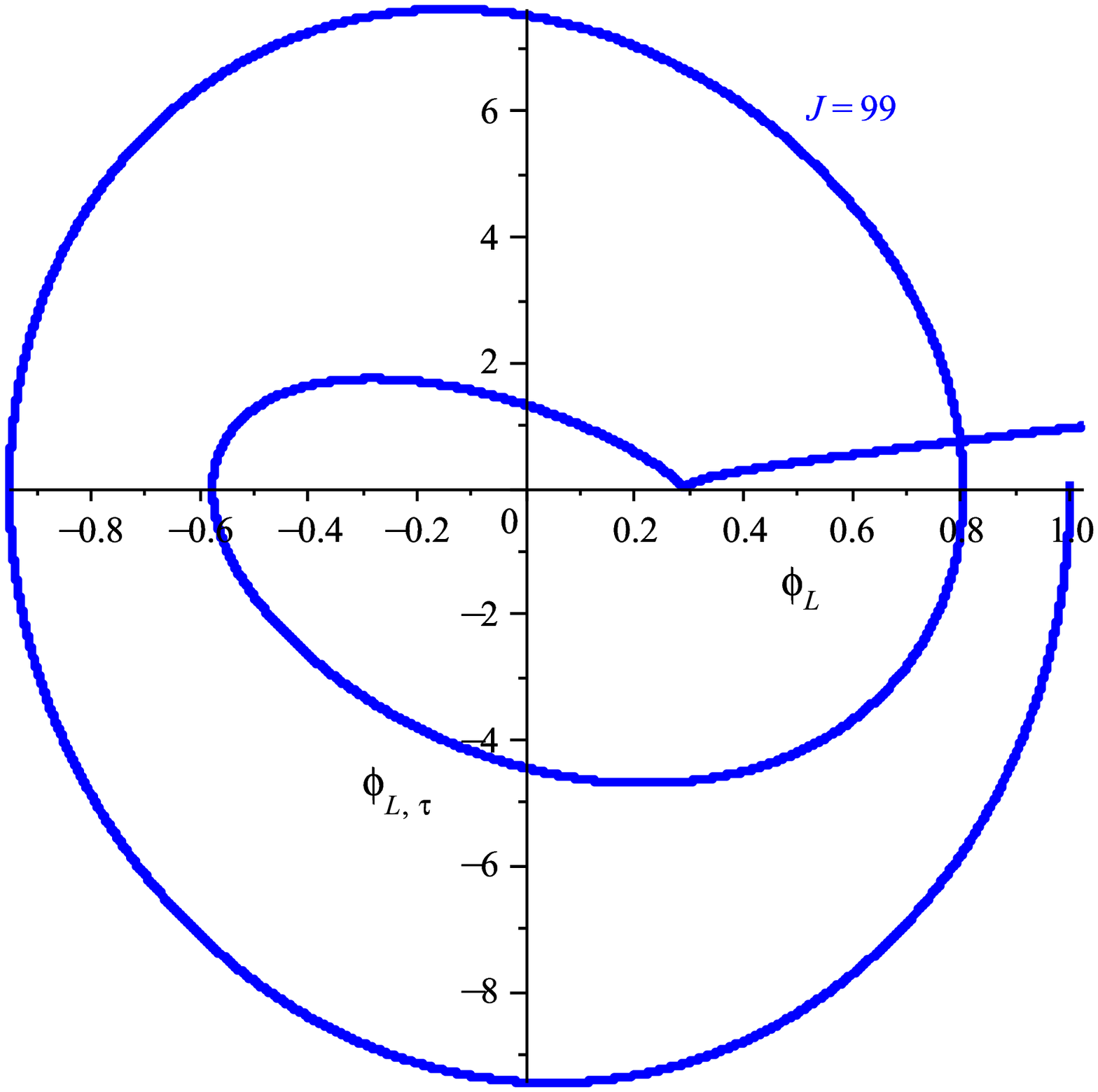}
\caption{Oscillatory behaviour exhibited from the value of the parameter $J=99$. 
The left picture is showed in the configuration space while the right picture 
was obtained on the corresponding phase space.}
\label{fig:3}
\end{figure}

For $p=3$ it was shown that lowest modes, $L=0$ and $L=1$, do not correspond 
to true perturbations but that they must be understood physically as a time 
translation and a spatial rotation of the unperturbed solution, 
respectively~\cite{vilenkin91}. These solutions are plotted in figure~(\ref{fig:1}). 
However, at cosmological level these lowest modes become significant in 
a inflationary scenario but not so with the other ones since they exhibit a 
damped oscillatory behaviour for short times~\cite{adams90}. We expect that 
for larger times the modes characterized by the oscillatory behaviour 
become important for the self-accelerated expansion of a late time brane-like 
universe~\cite{modified2012}, as seen in the figure~(\ref{fig:2}). 
Indeed, by considering even $J$ values, and for short times, we have perturbations 
provided with an oscillatory behaviour. As time increases, behaviour gets frozen 
at some finite value of the radius but subsequently restarts with some growing 
modes of oscillation which rapidly leading to instability thus ending up with highly 
nonspherical bubbles.
To explore the region of stability we can provide the representation of 
the solution for the perturbation from the point of view of the dynamical 
systems evolving in the phase space. For short times we observe that the 
trajectories for even values of $J$ tend to the point located at the origin 
which is a signal of stability. But such stability is deceptive. For the 
limit of larger time, it is observed that the trajectories are reset and these 
fastly move away from the origin what is a sign of an instability of the 
membrane~\cite{vilenkin91}.
In contrast, for odd values of $J$ it is noted from the beginning an oscillatory 
behaviour of the perturbation but as time passes the amplitude grows rapidly so 
that instabilities of the membrane can take place more quickly. 
Figure~(\ref{fig:3}) shows this last possibility of the behaviour for the 
solution $\phi_L$. With regards higher-dimensional de Sitter membranes 
we must mention that the behaviour is quite similar. 

\section{Concluding remarks}
\label{sec:6}

In this paper we reviewed the so-called Lovelock type brane gravity which, 
for even values of $p$ mimics the original Lovelock gravity terms while for odd values 
of $p$ resembles the necessary counterterms in order to have a well posed variational 
principle in gravity. Then, based on modern variational techniques, we have obtained
the criteria for which a Lovelock type extended object becomes extremal. We derived 
the general equations of motion written in a compact form by taking into account 
useful conserved brane tensors. Afterwards, we construct a covariant framework 
to study the conditions for which an extended object of this type becomes stable against 
infinitesimal perturbations.
Thence, we obtain the corresponding Jacobi equation which is manifested through 
a wave type equation governing the dynamics of the perturbations. To appreciate our 
framework, the special case of de Sitter membranes with spherical symmetry has been 
studied. In particular, the wave type equation is reduced to the usual Klein-Gordon equation, 
as in the standard DNG case so that we can take advantage of the analysis carried out in such a case.
Hence, we observe that for this geometry the analysis is insensitive to any model of 
the so-called Lovelock type brane gravity. This is by no means coincidental because the 
DNG model is part of this type Lovelock gravity as the first element of the 
expansion~(\ref{eq:Lbaction}) so it is expected that some of the properties already 
known for the DNG model reappear in the full theory. On the other hand, it should be mentioned also that the 
resulting Jacobi equation~(\ref{eq:ray1}) can be obtained from another viewpoint by 
considering a purely kinematical description of the normal deformations of the 
worldvolume~\cite{raycha95}. In such an approach, this equation corresponds to the 
so-called generalized Raychaudhuri equation which is an analogous to higher-dimensions 
of the Raychaudhuri equation describing point particles.
In this regard, we must mention that another interesting framework to analyse the stability
of brane universes was developed recently~\cite{vasilic2013}.
An even important question that we will investigate is whether we can extend our framework 
to the case of consider extended objects (or branes) as sources of gravity in the 
$N=(p+2)$-dimensional background spacetime. In such a case, we need to take careful in 
incorporating properly the Israel's junction conditions~\cite{davidson2006,charmousis2005,kofinas2014}.
In addition, it remains to explore the possible cosmological implications arising from this type 
of approach in the subject of brane cosmology. In particular, 
we can undertake a perturbation analysis of the so-called geodetic brane gravity which consists 
either of the first two or three terms of the Lovelock type brane gravity which has been 
reignited recently~\cite{davidson1,davidson2,ostrogradski,modified2012,davidson2005,qmodified2014,paston}.  
Another interesting issue to be explored is the existence of a critical dimension
for this type of gravity \cite{dadhich2010,kastor2012,dadhich2012,dadhich2016} as occur 
in the pure Lovelock gravity which could provide information on the solutions of the field equations. 
Investigation in these directions is presently underway. Finally, we hope that the 
perturbation analysis we have done here can provide a useful starting point to explore 
the issue of brane nucleation~\cite{davidson2005,vilenkin98,linde98}. 

\bigskip

\ack
MC and ER thank Aharon Davidson and Rub\'en Cordero for providing helpful 
discussions and comments on the geometric content of the work. 
MC would like to thank CONACyT in M\'exico for the support by the Grant: 
\textit{Repatriaciones, 2015, cuarta fase}. ER and CC thank ProdeP-M\'exico, 
CA-UV-320: Algebra, Geometr\'\i a y Gravitaci\'on. CC acknowledges partial 
support from CONACyT under grant CB-2012-01-177519-F. NB and ER thank
partial support from CONACyT under grant CB-2009-01-135297.
This work was partially supported by SNI (M\'exico).

\appendix

\section{A de Sitter membrane immersed in a Minkowski bulk}
\label{app1}

Consider the immersion of a $(p+1)$-dimensional de Sitter membrane embedded in a 
$N=(p+2)$-dim flat Minkowski spacetime in terms of the so-called Rindler 
coordinates~\cite{trodden2}
\begin{equation}
\fl
x^\mu = X^\mu (\tau, \theta_i) = 
\left(
\begin{array}{l}
H^{-1} \sinh (H \tau)
\\
H^{-1}\cosh (H \tau) \cos \theta_1
\\
H^{-1}\cosh (H \tau) \sin \theta_1 \cos \theta_2
\\
\qquad \qquad \vdots
\\
H^{-1}\cosh (H \tau) \sin \theta_1 \sin \theta_2 \cdots \sin \theta_{p-1} \cos \theta_p
\\
H^{-1}\cosh (H \tau) \sin \theta_1 \sin \theta_2 \cdots \sin \theta_{p-1} \sin \theta_p
\end{array}
\right),
\label{eq1}
\end{equation}
where $i=1,2,\ldots,p$. The orthonormal basis is given by the tangent vectors $e^\mu{}_a 
= \partial_a X^\mu$ and a normal vector $n^\mu$.
The induced metric $g_{ab} = \eta_{\mu \nu} e^\mu {}_a e^\nu {}_b$
results
\begin{equation}
dS_{p+1} ^2 = - d\tau^2 + H^{-2} \cosh^2 ( H\tau ) \,d \Omega_{(p)} ^2 ,
\end{equation}
where $d \Omega_{(p)} ^2 = d\theta_1 ^2 + \sin^2 \theta_1 d\theta_2 ^2
+ \sin^2 \theta_1 \sin^2 \theta_2 d\theta_3 ^2 + \sin^2 \theta_1 \sin^2 \theta_2 
\cdots \sin^2 \theta_{p-1} d\theta_p$ is the metric on the 
$p$-sphere and $H$ is a constant. With regards the extrinsic curvature, $K_{ab} = - 
\eta_{\mu\nu} n^\mu D_a e^\nu {}_b$, the non-vanishing components are 
\begin{eqnarray} 
K_{\tau \tau} &=& -H,
\\
K_{\theta_1 \theta_1} &=& H^{-1} \cosh^2 (H \tau),
\\
K_{\theta_2 \theta_2} &=& H^{-1} \cosh^2 (H \tau) \sin^2 \theta_1,
\\
K_{\theta_3 \theta_3} &=& H^{-1} \cosh^2 (H \tau) \sin^2 \theta_1 \sin^2 \theta_2,
\\
&\vdots&
\nonumber
\\
K_{\theta_p \theta_p} &=& H^{-1} \cosh^2 (H \tau) \sin^2 \theta_1 \sin^2 \theta_2
\cdots \sin^2 \theta_{p-1},
\end{eqnarray}
or, in the useful fashion $K^\tau {}_\tau = K^{\theta_i}{}_{\theta_i} 
= H$. Note that in this particular geometry we have that 
\begin{equation}
 K_{ab} = H g_{ab}, \qquad \mbox{and} \qquad K^{ab} = H g^{ab}.
\label{eq:A8} 
\end{equation}
This last fact allows us to compute easily the mean extrinsic curvature
\begin{equation}
K = g^{ab} K_{ab} =  (p+1) H.
\end{equation}
Indeed, this is expected because a de Sitter spacetime is a maximally
symmetric spcetime and in such a case the extrinsic curvature is
proportional to the induced metric. Similarly, we obtain that
$K_{ab} K^{ab} = (p+1) H^2$.
The contracted Gauss-Codazzi relation, ${\cal R}_{ab}
= K K_{ab} - K_{ac}g^{cd}K_{db}$, is useful to obtain the
components of the worldvolume Ricci tensor
\begin{equation}
{\cal R}_{ab} = p H^2 g_{ab}, \qquad \mbox{and} \qquad {\cal R}^{ab}=
p H^2 g^{ab}.
\end{equation}
or, in the fashion ${\cal R}^\tau {}_\tau = 
{\cal R}^{\theta_i}{}_{\theta_i} 
= pH^2$.
In consequence
\begin{equation}
\label{eq:Rab}
{\cal R} = g^{ab} {\cal R}_{ab} = K^2 - K_{ab} K^{ab} = p (p+1) H^2.
\end{equation}

\noindent
Furthermore, for this geometry the LBI and LBT are given in the compact form
\begin{equation}
J^{ab} _{(n)} = C_{(p,n)} H^n g^{ab}, \qquad \mbox{and} \qquad
L_n = C_{(p+1,n)} H^n,
\label{eq:Jab2}
\end{equation}
where we have introduced the notation $C_{(p,n)} = \Gamma (p+1) / 
\Gamma (p-n+1)$ where $\Gamma (n)$ is the Gamma function.

\section*{References}

\end{document}